\documentclass[aps,prl,twocolumn,groupedaddress,showpacs]{revtex4}
\usepackage{graphicx}
\usepackage{color}

\begin{document}

\title{Stroboscopic back-action evasion in a dense alkali-metal vapor.}

\author{G. Vasilakis, V. Shah, and M.V. Romalis}
\affiliation{Department of Physics, Princeton University, Princeton,
New Jersey, 08544, USA}

\date{\today}

\begin{abstract}

 We explore experimentally quantum non-demolition (QND) measurements of atomic spin in a hot
 potassium vapor in the presence of spin-exchange relaxation. We demonstrate a
 new technique for back-action evasion by stroboscopic modulation
 of the probe light. With this technique we study spin noise as a function of
 polarization for atoms with spin greater than 1/2 and
 obtain good agreement with a simple theoretical model. We point
 that in a system with fast spin-exchange, where the spin relaxation rate is changing with time, it is possible to
 improve the long-term sensitivity of atomic magnetometry by using QND measurements.

\end{abstract}


\pacs{42.50.Lc, 42.50.Ct, 03.65.Yz, 07.55.Ge}

\maketitle

Quantum non-demolition (QND) measurements form the basis of many
quantum metrology schemes \cite{PolzikRMP,Grangier,BraginskyRMP}. A
QND measurement can drive the system into a squeezed state
conditioned on the measurement result. In this state the uncertainty
of the measured variable is reduced below the standard quantum limit
(SQL) at the expense of an increase in the uncertainty of the
conjugate variable. A key ingredient in QND measurements is a
back-action evasion mechanism that decouples the measured variable
from the quantum noise of the probe field.

Here we explore a new back-action evasion scheme in a dense alkali
metal vapor in a finite magnetic field. A QND measurement of an
atomic spin component can be made by Faraday paramagnetic rotation
of off-resonant probe light \cite{TakahashiQND}. By stroboscopically
pulsing the probe light at twice the frequency of Larmor spin
precession, we achieve back-action evasion on one of the spin
components in the rotating frame,  while directing the quantum noise
of the probe beam to the conjugate rotating component. The
stroboscopic modulation of the probe was first suggested in the
context of mechanical oscillators \cite{CavesMechOsc}. In atomic
systems with non-zero Larmor frequency only more complicated schemes
involving two oppositely polarized vapor cells have been realized to
achieve back-action evasion \cite{PolzikNat}.

The QND measurements in a dense alkali-metal vapor allow us to study
atomic spin noise in the presence of various relaxation mechanisms.
The behavior of collective spin in the presence of decoherence is
not trivial \cite{KominisPRL,Geremia1,Geremia2}. We quantitatively
measure spin noise as a function of atomic polarization for K atoms
($I=3/2$) with spin-exchange, light scattering, and spatial
diffusion as the dominant sources of relaxation and obtain good
agreement with a simple model for quantum fluctuations.

Although QND measurements have been shown to increase the
measurement bandwidth without loss of sensitivity
\cite{QNDVishal,MitchelPRL}, it has been known for some time that
spin squeezing in the presence of a constant decoherence rate does
not significantly improve long-term measurement sensitivity
\cite{Huelga,AuzQND}. We point out that spin-exchange collisions,
which are the dominant source of relaxation in a dense alkali vapor,
cause non-linear evolution of the atomic density matrix with a
relaxation rate that changes in time. Under these conditions we show
theoretically that QND measurements can, in fact, improve the
long-term sensitivity of atomic magnetometers.

\begin{figure}
\centering
\includegraphics[width=8cm]{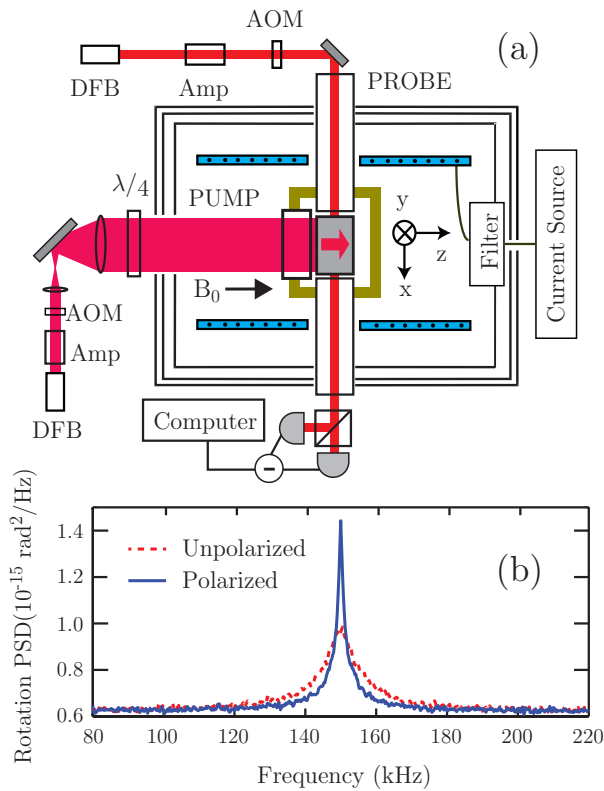}
\caption{(color online) (a) Experimental apparatus for QND RF
magnetometer. (b) Measured PSD for unpolarized (dashed) and highly
polarized atoms (solid). Each curve is the average of 1000
repetitions. Both curves were taken with the same probe intensity
and 10\% duty cycle. The atomic density was
$10^{14}$~cm$^{-3}$.}{\label{fig:SetupPSD}}
\end{figure}

The experimental setup is shown in Fig.~\ref{fig:SetupPSD}. The
atomic vapor is contained in a cylindrical, D-shaped  glass cell,
orientated in such a way that the probe beam goes through the long,
55 mm in length, dimension. We use a mixture of potassium in natural
abundance, 50 Torr of N$_2$ buffer gas for quenching and 400 Torr of
$^4$He to slow down the diffusion of alkali atoms. The cell is
heated in an oven with flowing hot air, and is placed inside a
double layer $\mu$-metal and a single-layer aluminum shield. A low
noise current source generates a homogeneous DC magnetic field in
the $\hat{z}$-direction, corresponding to a Larmor frequency of 150
kHz for K atoms. First order gradients of this field along the
direction of the probe beam are canceled with the use of a gradient
coil. In order to suppress current source noise and noise pickup of
the cables, passive low pass filters are placed inside the shields.
Narrow linewidth, amplified DFB lasers for the pump and probe beam
are used, and acousto-optic modulators provide fast amplitude
modulation of the light. The circularly polarized pump beam creates
atomic orientation in the $\hat{z}$-direction. It is turned off
after $10$~msec of pumping before probe measurements. The profile of
the pump beam is shaped using spherical abberation effects so that
the intensity is slightly higher at the edges of cell, where the
pumping requirements are higher due to the larger spin-destruction
rate from the wall relaxation. A linearly polarized probe beam far
detuned from the D1 line of K ($\lambda_{pr} \simeq 770.890$ nm) and
propagating along the $\hat{x}$-direction experiences Faraday
paramagnetic rotation which is measured with balanced polarimetry.
The signal is digitized with a fast, low noise A/D card and recorded
with a computer.

The back-action of the probe originates from the AC Stark shift
caused by quantum fluctuations of the circular polarization of the
light. For the conditions of our experiment, with large detuning,
high buffer gas pressure, and large optical density, the tensor
polarizability has a negligible contribution
\cite{HapperPRA,MitchelDec}. Then, the light shift noise can be
effectively described by a stochastic magnetic field along the
direction of the probe beam. During a short measurement of $F_x$ by
the probe beam this stochastic magnetic field rotates $F_z$
polarization into the $F_y$ direction, thus ensuring that the
product $\Delta F_x \Delta F_y$ satisfies the quantum uncertainty
relationship. In the presence of a DC magnetic field in the
$\hat{z}$-direction, the $x$ and $y$ components of the collective
spin undergo Larmor precession, so that over timescales larger than
the Larmor period both $F_x$ and $F_y$ accumulate the back-action
noise. The effect of back-action on the $F_x$ measurement in the
rotating frame can be suppressed using stroboscopic probe light that
turns on and off at twice the Larmor frequency. This way a
measurement is performed only when the squeezed distribution is
aligned with the probe axis in the laboratory frame.

The power spectral density (PSD) of a 3.6 msec recording of the
polarimeter output is shown in Fig.~\ref{fig:SetupPSD} for both
unpolarized and highly polarized atoms. The longitudinal spin
polarization does not change significantly on this time scale.  The
PSD can be described by a sum of a constant photon shot noise (PSN)
background and a Lorentzian-like atomic noise contribution
\cite{QNDVishal}. The deviation from the Lorentzian profile is
notable in our experiment due to the effect of diffusion in and out
of the probe beam (beam waist $\sim 220$~$\mu$m). As the atoms
diffuse through the probe beam, the measured collective spin
undergoes a random walk with correlation time characteristic of the
diffusion timescale. Note that for a coherent spin excitation with
an RF field diffusion through the probe beam does not lead to
decoherence. This is a manifestation of the general characteristic
that entangled states are more fragile. As discussed in
\cite{QNDVishal}, the shape of the atomic noise peak does not
influence the total optical rotation noise $\text{var} \left[
\phi_{at} \right]$, given by the area under the PSD curve. For
unpolarized atoms this noise area is a good measure of fundamental
atomic shot noise (ASN), since it is not affected by light-shift or
stray magnetic field noise, and the scattering of photons has an
insignificant effect on the quantum noise properties
\cite{MitchelPRL}. It provides a good reference for the
characterization of the atomic spin noise with polarized atoms. In
the fully polarized ensemble the spin-exchange collisions between
alkali atoms do not contribute to spin relaxation \cite{RFPaper},
and the spin noise linewidth is much smaller, as can be seen in
Fig.~\ref{fig:SetupPSD}.

\begin{figure}
\centering
\includegraphics[width=8cm]{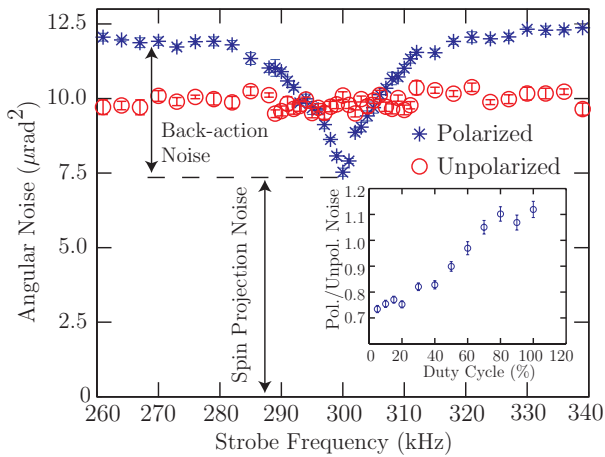}
\caption{(color online) Measurement of spin variance for unpolarized
and polarized ($P\approx 85\%$) atomic ensembles as a function of
stroboscopic frequency. The Larmor frequency is 150~kHz. While for
the unpolarized case the noise does not depend on the frequency, for
polarized atoms extra light shift noise appears at detunings from
the resonant condition. The data were taken using a probe with 10\%
duty cycle. Inset: Ratio of polarized to unpolarized noise (not
including PSN) as a function of the duty cycle of the strobe light.
All data points were acquired with the same average
intensity.}{\label{fig:SetupFrequencyDC}}
\end{figure}

The back-action evasion of the stroboscopic measurement is
demonstrated in Fig.~\ref{fig:SetupFrequencyDC}. The atomic noise is
evaluated by numerical integration of the measured PSD after
subtracting the constant PSN background. For polarized atoms, as the
strobe frequency departs from the resonance condition of twice the
Larmor frequency, light-shift noise is added to the ASN, and the
total noise increases until it reaches a maximum plateau. The
difference of the maximum and minimum values is a measure of the
back-action noise of the probe. In the case of unpolarized atoms,
there is no contribution of light-shift to the total noise, which
remains independent of the probe modulation frequency at the value
of the ASN. The back-action evasion is also observed when the noise
is plotted as a function of the duty cycle of the stroboscopic
probe. In the inset of Fig.~\ref{fig:SetupFrequencyDC} we normalize
each point by the corresponding unpolarized ASN and show that the
light-shift suppression is stronger for small duty cycle probe
pulses.

\begin{figure}
\centering
\includegraphics[width=8cm]{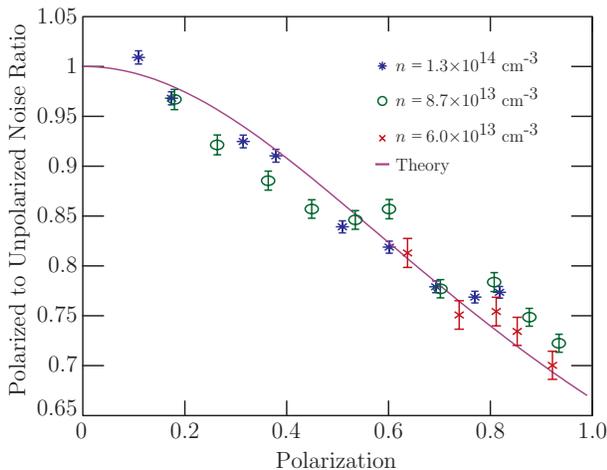}
\caption{(color online) Ratio of polarized to unpolarized ASN
(variance) as function of the mean longitudinal polarization of the
ensemble for three different densities. The duty cycle of the probe
was 10\%.}{\label{fig:PSDPolarization}}
\end{figure}

In Fig.~\ref{fig:PSDPolarization} the noise ratio for (partially)
polarized to unpolarized atomic ensembles is plotted as a function
of the longitudinal polarization for three different densities. The
ensemble polarization is found from the optical rotation induced in
the probe beam due to a known, small magnetic field in the probe
direction (B$_x \ll \text{B}_z$), slowly ac modulated to allow for a
lock-in detection of the signal. The largest uncertainty in this
measurement originates from the determination of the atomic density.
For this, we measure the coherent RF resonance curve at low
polarization and associate the measured linewidth with the
spin-exchange rate between alkali atoms \cite{RapidSpinExchange}. At
large values the ensemble polarization can also be directly
estimated from the transverse relaxation rate \cite{RFPaper}. The
two measurements give similar results for low atomic density, but
differ by $~10\%$ at the highest density. We believe this
discrepancy results from a nonuniform polarization profile of the
atomic ensemble, which becomes more pronounced at high densities due
to limited pumping power. Using gradient imaging we have measured
and minimized the polarization non-uniformity of the vapor.

The measured noise ratio can be described well with a simple
theoretical model. For the conditions of our experiment the density
matrix can be approximated for arbitrary longitudinal polarization
$P$ by the spin temperature distribution \cite{AppeltPumping}:
$\rho=e^{\beta F_z}/Z$, where $Z$ is the partition function and
$\beta=\ln\left[(1+P)/(1-P)\right]$. Then, taking into account the
two hyperfine manifolds of the alkali-metal atoms \cite{QNDVishal},
the ASN variance of the collective spin composed of $N_a$ atoms can
be written:
\begin{eqnarray}\label{eq:SpinNoise}
& &\langle  F_x^2 \rangle = \sum_{m=-a}^{a} \frac{e^{\beta m} \left[ a(a+1)-m^2 \right]}{2 N_a Z}+ \nonumber\\
& &  \sum_{m=-b}^{b} \frac{e^{\beta m} \left[ b(b+1)-m^2 \right]}{2
N_a Z}
\end{eqnarray}
Here, $a=I+1/2$ and $b=I-1/2$, with $I$ being the nuclear spin. One
can see that in contrast to a spin-1/2 system, for $I=3/2$ the ASN
power is smaller for polarized atoms by a factor of 2/3
 compared with unpolarized atoms, in agreement with the experiment. These data address some of the
issues raised in \cite{Geremia1} regarding collective measurements
on partially polarized atomic states. They also disprove the claim
in \cite{KominisPRL} that correlated spin relaxation due to
spin-exchange collisions does not lead to atomic noise.

\begin{figure}
\centering
\includegraphics[width=8cm]{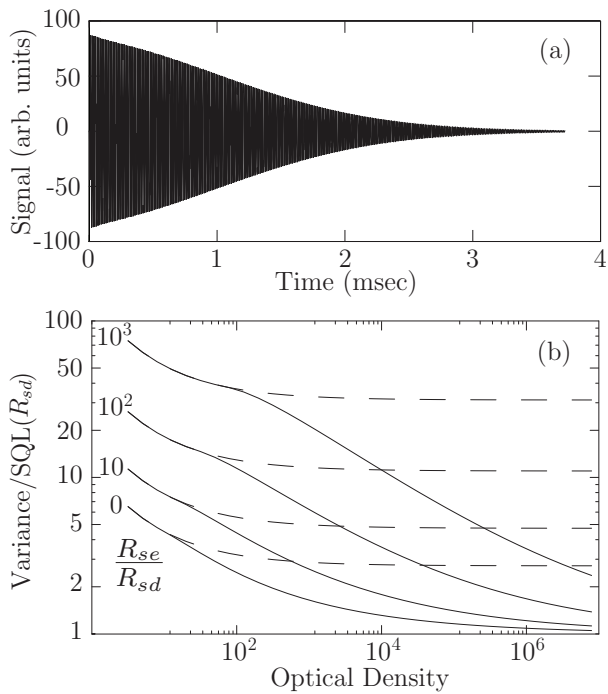}
\caption{ (a) Experimental measurement of $F_x$ at high density
($n\approx 6 \times 10^{13}$~cm$^{-3}$) following a short magnetic
field pulse, showing changes in the transverse relaxation rate. For
this data the probe beam scattering rate was increased. (b)
Calculated variance in the estimate of the magnetic field relative
to SQL as a function of the optical density for various
spin-exchange rates. Dashed lines -- single-pulse measurement,
solid-lines -- two pulse measurement with spin-squeezing.
}\label{fig:Nonlinear}
\end{figure}

As can be seen in Fig.~\ref{fig:SetupPSD}, the resonance linewidth
is significantly reduced for high spin polarization due to
suppression of the spin-exchange relaxation. In the time domain this
is manifested by a non-exponential decay of the transverse spin
polarization, shown in Fig.~\ref{fig:Nonlinear}(a). In a highly
polarized vapor the initial spin relaxation rate is suppressed. This
allows one to improve the overall long-term measurement sensitivity
using QND measurements.

To model this behavior quantitatively we consider a measurement
scheme using two short pulses of probe light \cite{PolzikHF}. The
first pulse is applied immediately after turn-off of the pump beam
and the second after a measurement time $t_m$. The best measurement
of the magnetic field is obtained using an estimate $S_x(t_m)-S_x(0)
{\rm cov}[S_x(0),S_x(t_m)]/{\rm var}[S_x(0)]$, where $S_x(0)$ and
$S_x(t_m)$ are measurements of spin projection from the two probe
pulses. For simplicity we consider a spin-1/2 system here. One can
show that ${\rm var}[S_x]=(1+1/\epsilon {\rm OD})N_A/4 $, where
$\epsilon$ is the strength of a far-detuned probe pulse, given by
the product of pulse duration and photon scattering rate, ${\rm OD}$
is the optical density on resonance, and $N_A$ is the number of
atoms. The covariance of the two measurements is given by (for
$t_m>0$) \cite{Gardiner}
\begin{equation}
{\rm cov}[S_x(0),S_x(t_m)]= (N_A/4) \exp[- \int \nolimits _{0}^{t_m}
R(t') dt'],
\end{equation}
where $R(t)$ is a time-dependent transverse spin-relaxation rate. In
the presence of spin-exchange collisions the relaxation rate can be
approximated by $R(t)=R_{sd}+(1-P_z) R_{se}$ \cite{RFPaper}.  Using
this model we optimize the measurement procedure with respect to the
strength of first and second probe pulses and $t_m$. We assume that
the initial state preparation time is negligible and the measurement
repetition time is equal to $t_m$. The results of the model are
plotted in Fig.~\ref{fig:Nonlinear} for varying spin-exchange rates.
For comparison, we also plot the variance of a single-pulse
measurement after time $t_m$, which does not rely on spin-squeezing.
The results are scaled relative to the SQL limit for $N_A$ atoms
with spin relaxation rate $R_{sd}$, $\delta B^2_{SQL}=2R_{sd}/(N_A t
\gamma^2)$, where $t$ is the total measurement time and $\gamma$ is
the gyromagnetic ratio.

It is instructive to compare our results with those of
\cite{Huelga}. In the absence of spin-squeezing and spin-exchange
relaxation, the smallest possible magnetic field variance is given
by $ e \delta B^2_{SQL}$, in agreement with \cite{Huelga}. Using the
two-pulse measurement one can reduce the variance by a factor of
$e$, the same factor as obtained in \cite{Huelga} with partially
entangled states. In the presence of spin-exchange relaxation, the
sensitivity is degraded for the one-pulse scheme, but reaches the
same $\delta B^2_{SQL}$ using two pulses. Therefore, QND techniques
can eliminate the effects of spin-exchange relaxation, but cannot
significantly exceed the sensitivity corresponding to a constant
relaxation rate. These results also apply to hyperfine transitions
which are broadened by spin-exchange \cite{HapperCPT}, and, more
generally, to other relaxation effects due to non-linear
interactions, such as solid-state dipolar spin coupling
\cite{Abragham}.

In summary, we have explored quantum non-demolition measurements of
collective spin in a dense alkali-metal vapor. We demonstrated a new
stroboscopic technique for back-action evasion and used it to
measure atomic spin noise as a function of spin polarization in the
presence of several spin-relaxation mechanisms. We considered QND
measurements in a system with non-linear spin relaxation and showed
theoretically that they can improve the long-term sensitivity in
atomic spectroscopy. This work was supported by NSF and ONR MURI.

\bibliography{Bibliography}

\end{document}